\documentstyle[prl,twocolumn,aps,psfig,epsf]{revtex}

\begin{document}
\draft
\title{Odd triplet superconductivity in superconductor/ferromagnet multilayered
structures.}
\author{A.F. Volkov$^{1,2},$F.S.Bergeret$^{1}$, and K.B.Efetov$^{1,3} $}
\address{$^{(1)}$ Ruhr-University Bochum, D-44780 Bochum, Germany\\
$^{(2)}$Institute of Radioengineering and Electronics of the Russian \\
Academy of Sciences, Moscow 103907, Russia\\
$^{(3)}$L.D. Landau Institute for Theoretical Physics, 117940 Moscow, Russia }

\maketitle

\begin{abstract}
We demonstrate that in multilayered superconductor-ferromagnet structures a
non-collinear alignment of the magnetizations of different
ferromagnetic layers generates a triplet superconducting condensate which is
odd in frequency. This triplet condensate coexists in the superconductors
with the conventional singlet one but decays very slowly in the ferromagnet,
which should lead to a large Josephson effect between the superconductors
separated by the ferromagnet. Depending on the mutual direction of the
ferromagnetic moments the Josephson coupling can be both of $0$ and $\pi $
type.
\end{abstract}
It is well known that ferromagnetism and superconductivity are antagonistic
phenomena; ferromagnetism strongly suppresses superconductivity. { This
suppression in  superconductor/ferromagnet (S/F) layered structures  is  caused mainly by  the exchange interaction. This means in  particular that the singlet Cooper pairs cannot penetrate into the F layers over a noticeable length, since the exchange energy is 
by orders of magnitude    larger than  the Cooper
binding energy. 
 Thus, the singlet pairs are destroyed by the exchange field because the
spins of the electrons cannot be antiparallel anymore.}
{\ This suppression of superconductivity can be reduced in S/F structures
if the ferromagnetic layers are ultrathin, but, generally, the destruction
of the singlet superconductivity by a homogeneous exchange field can hardly
be avoided.}

The situation may be different if the spins of the superconducting pairs are
parallel to each other. It is clear that such a triplet superconductivity is
not sensitive to the exchange field and the coexistence of the
superconductivity and ferromagnetism becomes possible. Unfortunately, the
triplet pairing is a rather exotic phenomenon and has been
observed until now only in superfluid He$^{3}$ and in a superconducting
material Sr$_{2}$RuO$_{4}$ \cite{SrRuO}. It is expected to be very sensitive
to disorder \cite{SrRuO,Larkin}, which makes its observation even
more difficult. In order to satisfy fermionic commutation relations the
condensate function, even in frequency, should be odd in the momentum of the pair and this is the reason why it is so sensitive to impurities. Another possibility was suggested by Berezinskii long ago \cite{Berez}. He
conjectured that the triplet superconductivity might be possible if the
condensate function were even in the momentum but odd in the frequency.
Attempts to find conditions for the existence of such an odd
superconductivity were done in several works much later \cite{Balats} but
the results were not encouraging (in Ref. \cite{Balats} a singlet pairing
odd in frequency and in the momentum was considered).

In this Letter, we demonstrate that the odd triplet superconductivity is not
exotic at all and is ${\em inavoidable}$ in  multilayered $S/F$
structures  with  conventional superconductors if the
directions of the magnetization of the  different F-layers are neither parallel
nor antiparallel to each other. In this case the triplet condensate (TC) can easily penetrate into the ferromagnetic layers over long distances and result in
supercurrents through the ferromagnets. The conditions for the realization
of the odd triplet superconductivity do not seem to be problematic from the
experimental point of view and we hope that proper measurements will be done
in the nearest future.

\begin{figure}
\epsfysize = 4.6cm
\vspace{0.2cm}
\centerline{\epsfbox{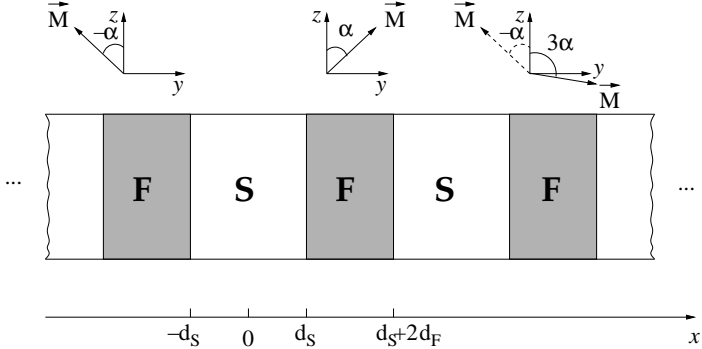}}
\vspace{0.2cm}
\caption{\label{Fig.1}Schematic diagram of a S/F multilayered structure. On top: the magnetization ${\bf M}$ in different F layers. The ${\bf M}$ vector drawn by solid lines correspond to chiralities of the same sign. Opposite chirality is obtained  if, for example, the magnetization of the right F layer is rotated in the opposite direction (dashed line).}
\end{figure}

A generation of a triplet condensate by a non-homogeneous magnetization
(domain walls) and its penetration into a ferromagnet has been discussed
recently \cite{BVE2,Swed}.  In the structures considered in these works only changes in the conductivity of the system were analyzed. However, the  detection of the triplet component in such structures is a quite difficult task. In contrast,
we predict in this Letter a new type of {\it superconductivity} (odd triplet) in S/F structures and discuss how to  identify it experimentally. To be specific, we consider a $S/F$ multilayered structure (Fig. \ref{Fig.1}) in which  the new
superconducting state might be observed. In this state, the superconductivity in the $S$
regions is caused by the singlet component (SC) and takes place in the plane of the layers, while the transverse superconductivity through
the $F$ layers is mainly due to the TC. Moreover, the relation between the
condensate current $I_{S}$ and a phase difference $\varphi $ depends  in a
crucial way on chirality of the magnetic moment ${\bf M}$ in space.

The multilayered $S/F${\it \ }structure we consider is represented in
Fig.\ref{Fig.1}. The (in-plane) magnetizations ${\bf M}$ of the neighboring $F$ layers are not parallel to each other and the angle between them is $2\alpha $. {\  In order  to achieve such a non-collinear alignment one can employ an  exchange-biased spin-valve or ferromagnets with strong anisotropy and different easy-axis of magnetization.}
 It will be shown
that in such a structure the TC may arise if the thickness of the
superconducting layers $2d_{S}$ is less or comparable with the
coherence length $\xi _{S}=\sqrt{D_{S}/2\pi T}.$ The TC penetrates into the
F layer over a long distance $\xi _{T}=\sqrt{D_F/2\pi T}$ and ensures a Josephson coupling
between the nearest $S$ layers. In the case under consideration the relation
between the condensate current $I_{S}$ and the phase difference $\varphi $ has the conventional form $%
I_{S}=I_{c}\sin $ $\varphi .$ However, the sign of the critical current $%
I_{c}$ depends on the chirality, namely, it is positive if the rotation
angle $2\alpha _{i}$ of the magnetization at the $S_{i}$ layer has the same
sign for neighboring $S$ layers ($S_{i}$ and $S_{i+1}$) and it is negative
if the signs of the  rotation are opposite for neighboring $S$ layers ($%
\alpha _{i}\alpha _{i+1}<0$).
Now, let us explain how these results were obtained. 
We consider the
simplest case of a dirty system for which the Usadel equations can be applied to
calculate the condensate function $f,$ i.e., we assume that the condition $%
\Delta <J<\tau ^{-1}$ is fulfilled, where $\Delta $ is the order parameter
in $S$ (in the ferromagnet $\Delta =0$), $J$ is the exchange energy and $%
\tau $ is the momentum relaxation time. However, even in this case the
problem remains too complicated and we need to make further assumptions.
First, we assume that the thickness of the $F$ layers $2d_{F}$\ exceeds the
length $\xi _{T}$. In this case the Josephson coupling is determined by an overlap of
exponentially decaying condensate functions induced by the nearest $S$
layers. {\ We also assume that the temperature is close to the critical temperature $T_c^*$ of the structure ($|T-T_c^*|/T_c^*\ll1$).} In this case the Usadel equation can be linearized and what
remains to do is to find the condensate function $f\left( x\right) $ \ by
solving the linearized Usadel equation   in the $S$ and $F$ regions.
The
Usadel equation  for the matrix condensate function $\check{f}$ takes the form \cite{Usad,Joseph}
\begin{equation}
D_{S}\partial _{xx}^{2}\check{f}-2|\omega |\check{f}=2i\check{\Delta} \text{, \ \ S layer}
\label{UsadS}
\end{equation}

\begin{eqnarray}
\lefteqn{D_{F}\partial _{xx}^{2}\check{f}-2|\omega |\check{f}+iJ{\rm sgn}\omega \left\{ [\hat{\sigma} _{3},\check{f}]_{+}\cos
\alpha +\right.} \nonumber  \\
&\!\!\! \left.+\hat{\tau} _{3}[\hat{\sigma} _{2},\check{f}]_{-}\sin \alpha \right\} =0\text{, \ \ F layer}
\label{UsadF}
\end{eqnarray}
Here $D_{S,F}$ are the diffusion coefficients in the $S$ and $F$ layers
respectively, $\check{\Delta} =i\hat{\tau} _{2}\otimes \hat{\sigma} _{3}\Delta $, $\omega =\pi T(2n+1)$ is the Matsubara frequency$,$ $J$ is the
exchange energy (strictly speaking, ${\bf J}$ is a vector directed along the
magnetization vector ${\bf M}$), $\hat{\tau} _{i}$ and $\hat{\sigma} _{i}$ are the Pauli
matrices in the Nambu and spin space, respectively. The brackets $[...]_{\pm
}$ denote an  anti-commutator and a commutator. We assume that the ${\bf M}$ vector
lies in the $(y,z)$ plane and $\alpha $ is the angle between the $z-$axis
and ${\bf M}$. 
Eqs.(\ref{UsadS}-\ref{UsadF}) have to be
supplemented by boundary conditions. In the case of a perfect $S/F$
interface (the reflection coefficient is very small) they have the form \cite{Zaitsev}
\begin{eqnarray}
\check{f}(d_{S}+0)-\check{f}(d_{S}-0) &=&0,  \label{BoundCond} \\
\gamma \partial _{x}\check{f}(d_{S}+0)-\partial _{x}\check{f}(d_{S}-0) &=&0\; ,  \nonumber
\end{eqnarray}
where $\gamma =\sigma _{F}/\sigma _{S},$ $\sigma _{F,S}$ are the
conductivities in the $F$ and $S$ regions (for simplicity we do not take
into account a dependence of the conductivity $\sigma _{F}$ on the spin
directions). {\ If the SF interface resistance $R_b$ is finite the r.h.s of the first equation equals $\gamma_b\xi_J\partial_x\check{f}(ds+0)$, where $\gamma_b$ is proportional to $R_b$ \cite{Tc}. For simplicity we set $\gamma_b=0$. A generalization of the results for the case of a finite $\gamma_b$  does not lead to qualitative changes.}

\begin{figure}
\epsfysize = 4.7cm
\vspace{0.2cm}
\centerline{\epsfbox{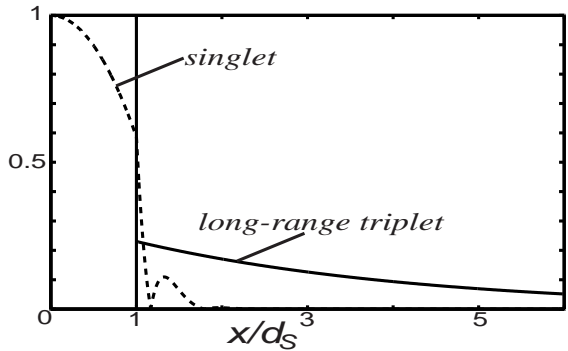}}
\vspace{0.2cm}
\caption{\label{Fig.2} The normalized absolute value of the SC (dashed line) in the S and the F layers. The solid line represents the absolute value of the TC in the F layer. Note that the SC (TC) is an even (odd) function of $x$. Here $d_S/\xi_T=d_S/\xi_S=0.3$, $d_S/\xi_J=5$, $\gamma=0.1$ and $\sin\alpha=0.5$. { We have assumed a finite SF interface resistance and set $\gamma_b=0.5$.}}
\end{figure}

The condensate matrix function $\check{f}$ has following form
\begin{equation}
\check{f}\left( x\right) =i\left( \hat{f}_{1}\left( x\right) \hat{\tau} _{1}+\hat{f}_{2}\left(
x\right) \hat{\tau} _{2}\right)\; ,  \label{CondFunc}
\end{equation}
where $\hat{f}_{1,2}$ are matrices in the spin space. In the $F$ layers they can
be written as
\begin{eqnarray}
\hat{f}_{1} &=&B_{1}\left( x\right) \hat{\sigma} _{1}+B_{2}\left( x\right) \hat{\sigma} _{2},%
\text{ }  \label{f1,2} \\
\hat{f}_{2} &=&B_{0}\left( x\right) +B_{3}\left( x\right) \hat{\sigma} _{3}\;.
\nonumber
\end{eqnarray}
In the $S$ layers the functions $\hat{f}_{1,2}$ have the same form, but the
coefficients $B$ should be replaced by $A$.  The functions $%
B_{i}\left( x\right) $ have the form: $B_{i}(x)\sim b_{i}\exp (\mp \kappa (x\mp d_{S})).$ Substituting
these expressions for $B_{i}(x)$ into Eqs. (\ref{UsadS}-\ref{UsadF}) {\ one obtains a set of algebraic equations that determine the eigenvalues $\kappa$.}
In
the $S$ region all condensate components (SC and TC) are decoupled and
there is only one solution for $\kappa $: $\kappa ^{2}=2|\omega |/D_{S}.$
In the $F$ region the SC and TC are coupled by the exchange interaction (if $%
\alpha $ is not zero) and there are three eigenvalues for $\kappa .$ Two of
them are equal to $\kappa _{\pm }=\kappa _{J}(1\pm i)$, and determine a fast
decay of the condensate in the $F$ layers ( dashed line in Fig. \ref{Fig.2}),{\ here $\kappa_J=\sqrt{J/D_F}$} .  This result demonstrates the well known short-range
penetration of the superconducting condensate into the ferromagnet (see for
example the review article \cite{Proshin}). However, the third solution for $\kappa$ is completely different and is given by 
$\kappa _{\omega }=\sqrt{2|\omega| /D_{F}}$.
 Thus, we see  that the huge exchange energy $J$ entering the
solution for $\kappa_{\pm }$ is replaced  by a small energy $%
\omega $ of the order of $T$. At large distances it is the solution with 
$\kappa _{\omega }$ that determines the penetration of the superconducting
condensate into the $F$ regions (see Fig. \ref{Fig.2}). It corresponds to the triplet component and,  as it has been discussed in the introduction, 
there is no wonder that the exchange field does not {\ influence it.}

Having determined the solutions for $\kappa$, we can write the solutions for $%
B_{i}\left( x\right) $. For example, the coefficient $B_{1}(x)$ may be
written at $x>d_{S}$ in the form
\begin{equation}
B_1(x)=\sum_i b_{1i}\exp\left[-\kappa_i(x-d_S)\right]\;, \label{B1}
\end{equation}
where $b_{1i}=b_{1\omega},b_{1\pm}$ and $\kappa_i=\kappa_{1\omega},\kappa_{1\pm}$.
The coefficients $b_{1\pm }$ are related to the coefficients $b_{3\pm
} $ that determine the magnitude of the SC by 
$b_{1\pm }=\mp \sin \alpha\, {\rm sgn}\omega \cdot b_{3\pm}$.
  
What remains to be done is to determine the coefficients $a_{i},b_{i}$ \
using the boundary conditions (\ref{BoundCond}) for an arbitrary angle $%
\alpha .$ In a general case the expressions for the coefficients are
 cumbersome, and for clarity we present here only results for small $\alpha .$ In
zero order in $\alpha $ only coefficients in the function $\hat{f}_{2}$ differ
from zero. For example, for $b_{3\pm }$ we obtain
\begin{equation}
b_{3\pm }=\frac{\Delta }{2i|\omega |}\frac{1}{1+\gamma \kappa _{\pm
}/(\kappa _{S}\tanh \theta _{S})}  \label{b3}
\end{equation}
where $\kappa _{S}=\sqrt{2|\omega| /D_{S}},$ $\theta _{S}=$ $\kappa
_{S}d_{S}.$ This solution corresponds to those obtained earlier  (see  \cite{Proshin} and references therein) where the
case of a parallel or antiparallel magnetization alignment was analyzed. It
is valid in the limiting cases $\theta _{S}<<1$ and $\theta _{S}>>1$ ( if $T$ is close to the critical temperature $T_c^*$).

If the magnetization vectors in the neighboring $F$ layers are inclined with
respect to each other by an angle $2\alpha ,$ the situation changes
qualitatively. First, the function $\hat{f}_{1}$ that describes the TC is no longer
zero   because the  terms proportional to $b_{1\pm}$ in the expression for $B_{1}(x)$ (Eq. (\ref{B1}))  are related to $b_{3\pm}$, and hence they are  finite in the main approximation. Secondly, in
order to fulfill the matching conditions (\ref{BoundCond}), one has to take
into account the first long-range term in Eq.(\ref{B1}). Using the boundary
conditions, we obtain for the coefficient $b_{1\omega }$ the following expression 
\begin{equation}
\!\!\!\!\!\!\!\!\! b_{1\omega}\!\!=\!-\frac{\Delta}{\omega}\sin\alpha\frac{\kappa_J\tanh\theta_s}{\cosh^{2}\theta_s|\gamma\kappa_+/\kappa_s+\tanh\theta_s|^2(\kappa_\omega\tanh\theta_s+\kappa_s/\gamma)}\; .
\label{b1}
\end{equation}
Contrary to the SC determined by the solution $B_{3}(x)$ and $A_3(x)$, the TC is an odd
function of $x$.  This is seen from Eqs. (\ref{B1}), (\ref{b1}) and the fact the $\alpha$ has different sign at $x>d_S$ and $x<-d_S$. {\ According to the boundary conditions (\ref{BoundCond}) (continuity of the condensate function and current conservation), the TC induced in the F layer penetrates into the superconductor.  The corresponding solution in the S layer has the form: $A_1(x)=a_1\sinh(\kappa_Sx)$.} In the  limit  $T<<J$, the TC penetrates into the 
$F$ layer over the length of the order $1/\kappa _{T}=\sqrt{D_{F}/2\pi T}$ which
is much greater than the SC penetration length $1/\kappa _{\pm }$. One can
see from Eq. (\ref{b1}) that the amplitude $b_{1\omega }$ of the long-range
TC is an odd function of the Matsubara frequency $\omega $ and is symmetric
in the momentum space as in the case of the TC which arises in a ferromagnet
with a non-homogeneous $M$ near a S/F interface \cite{BVE2}. This new type
of the condensate, odd in $\omega $ and even in the momenta $p$, has been
proposed by Berezinksii \cite{Berez} in order to explain the pairing
mechanism in He$^{3}$ (it was proven later that the condensate in He$^{3}$
is in fact even in $\omega $ and odd in $p$). The solution we present here
corresponds to this hypothetical pairing, which means that we have found a
concrete realization of the idea. It follows from the geometry of the
structure we consider that the odd triplet superconductivity we found exists
in the transverse direction.

It is seen from Eqs. (\ref{b3}) and (\ref{b1}) that the amplitudes of the SC and TC {\ at the S/F interface}
are comparable if $\xi _{J}<<\xi _{T}$ and $\alpha $ and $\theta _{S}$ are
of the order of $1$. If the latter condition is not satisfied (that is, the
thickness of the $S$ layer $2d_{S}$ is large in comparison with $%
\xi _{S})$, the TC decays exponentially in $S$ and its amplitude is small. {\ Therefore we 
 calculate the Josephson current  between the neighboring $S$ ($S_{1}$
and $S_{2}$) layers in the case $\theta_S\ll1$ and} assuming that the condition $\xi _{J}<<\xi _{T}<2d_{F}$
is fulfilled. Then, the Josephson current is due to the overlap of the TC
induced near each $S/F$ interface. In this case the TC in $F$ is described
by the expression

\begin{eqnarray}
\lefteqn{\check{f}_{trip}(x)=i\hat{\tau} _{1}\otimes \hat{\sigma} _{1}b_{1\omega }\exp (-\kappa _{\omega
}(x-d_{S}))+}\nonumber\\
&\;\;+\check{S}\cdot i\hat{\tau} _{1}\otimes \hat{\sigma} _{1}\cdot \check{S}^{+}\tilde{b}_{1\omega}\exp (\kappa _{\omega }(x-d_{S}-2d_{F}))  \label{fTr}
\end{eqnarray}
Here, the matrix of a gauge transformation $\check{S}=\cos (\varphi /2)+i\hat{\tau} _{3}\sin (\varphi /2)$ allows
us to take into account the phase difference $\varphi $ between the
neighboring $S$ layers (we assume that the phase of the $S_{1}$ layer is
zero). The coefficients have different signs ($b_{1\omega }=-\tilde{b}_{1\omega }$)
if the magnetization $M$ at both $S$ layers rotates in the same direction,
and $b_{1\omega }=\tilde{b}_{1\omega }$ in the opposite case (different signs of
chiralities). The condensate current $I_{S}$ between $S_{1}$ and $S_{2}$ is
given by the formula
\begin{equation}
I_{S}=(L_{y}L_{z})\sigma _{F}(\pi iT/4e)Tr(\hat{\tau} _{3}\otimes \hat{\sigma}
_{o})\sum_{\omega }\check{f}\partial _{x}\check{f}  \label{Current}
\end{equation}
From Eqs.(\ref{fTr}-\ref{Current}), we get $I_{S} =I_{c}\sin \varphi$, where 
\begin{equation}
I_{c}=-(2\pi T/eR_{F})2d_F\sum_{\omega
}\kappa _{\omega }b_{1\omega }\tilde{b}_{1\omega }\exp (-2d_{F}\kappa _{\omega })  \label{IS} 
\end{equation}
and $R_{F}=2d_F/((L_{y}L_{z})\sigma _{F})$ is the resistance of the $F$
layer in the normal state. Substituting Eq.(\ref{b1}) into Eq.(\ref{IS}), we
find for the critical current in the limit $\theta_s\ll1$  
\begin{equation}
\!\! eR_{F}I_{c}=\frac{4T}{\pi}\left(\frac{\Delta}{T}\right)^{2}\frac{(\kappa
_Jd_{S})^{2}(d_F\kappa_T)  e^{-2d_{F}\kappa _{T}} \sin \alpha _{1}\cdot \sin \alpha _{2} }{|\kappa _{So}d_{S}+\gamma \kappa _{J}(1+i)/\kappa_{So}|^{4}(\kappa
_{T}d_{S}+1/\gamma)^{2}}  \label{CrCurrent}
\end{equation}
where $\kappa _{So}=\sqrt{2\pi T/D_{S}}$.

If  the magnetization vector at each $S$ layer rotates in the same direction  ($\alpha _{1}=\alpha _{2}$), then the critical Josephson current $I_{c}$
is positive (0-contact). If  ${\bf M}$ rotates at $S_{1}$ and $S_{2}$ in  opposite directions ($\alpha _{1}=-\alpha _{2}$), then the
critical current $I_{c}$ is negative ($\pi -$contact). In the latter case a
phase difference $\pi $ is established between neighboring $S$ layers in a
multilayered $S/F$ structure. We would like to note that the mechanism of
the $\pi -$contact considered here is completely different from that
suggested in Ref.\cite{PiContact} and observed in Ref.\cite{Ryaz}. In our
case, the negative critical current is caused by the TC, but not by the SC
as in Ref. \cite{PiContact}, and, in addition, is realized only if the
chiralities at the $S_{1}$ and $S_{2}$ are different. The possibility of switching between the $0$ and $\pi$-states by changing the angle $\alpha$ may find applications in superconducting devices.

{\ The effect described above exists for all temperatures below  the critical temperature $T_{c}^{\ast }$ which  is determined from the self-consistency equation.} In
the main approximation in the small angle $\alpha $ it agrees with that
obtained in Refs.\cite{Tc},\cite{Proshin} for the case of the parallel
orientation of the magnetization. A correction to $T_{c}^{\ast }$ due to a
small misalignment of the magnetizations in neighboring $F$ layers, {\ i.e. due to the TC,} is
proportional to $\sin ^{2}\alpha$, but we are not interested in this
correction here. We note that the critical temperature\ $T_{c}^{\ast }$ for
the case of arbitrary $\alpha $ was analyzed in Ref.\cite{Buzd}. However,
the form of the condensate function presented in that work is not correct
because the authors started from an equation different from Eq. (\ref{UsadF}
) (instead of the commutator, they wrote the product $J\hat{\vec{\sigma}} \cdot \hat{f}$ ). As
a result, the long range triplet component was completely lost.
In conclusion, we have predicted a new type of superconductivity. It was demonstrated that in superconductor-ferromagnet
structures a non-collinear alignment of the exchange fields in the ferromagnetic
layers generates a triplet component of the superconducting condensate and
this component is odd in the frequency. The odd triplet condensate
penetrates into the ferromagnet over long distances and is not sensitive to
impurities. In the structure we suggest, a Josephson effect between
superconductors separated by the ferromagnet is possible and the critical
current can be measured. The Josephson contact can be both of $0$ and $\pi $%
-type depending on the arrangement of the magnetic moments. We hope that the
effects considered in this paper will be observed experimentally in the
nearest future.

We would like to thank SFB 491 and GIF for financial support.

\end{document}